\documentclass[a4paper]{article}

\usepackage{INTERSPEECH2022}

\usepackage{multicol,multirow}
\usepackage{cite}
\title{Leveraging Acoustic Contextual Representation by Audio-textual Cross-modal Learning for Conversational ASR}
\name{Kun Wei$^{1,2*}$, Yike Zhang$^{2*}$, Sining Sun$^2$, Lei Xie$^1\dag$, Long Ma$^2$}
\address{
  $^1$Audio, Speech and Language Processing Group (ASLP@NPU), School of Computer Science, Northwestern Polytechnical University, Xian, China\\
  $^2$Cloud Xiaowei, Tencent, Beijing, China
  \thanks{$^*$Equal contribution.
     }
    \thanks{$^{\dag}$Corresponding author.}}
\email{ethanwei@mail.nwpu.edu.cn, lxie@nwpu-aslp.org, \{yikezhang,siningsun,malonma\}@tencent.com}

\begin{document}

\maketitle
\begin{abstract}
  Leveraging context information is an intuitive idea to improve performance on conversational automatic speech recognition (ASR).
  Previous works usually adopt recognized hypotheses of historical utterances as preceding context, 
  which may bias the current recognized hypothesis due to the inevitable historical recognition errors.
  To avoid this problem,
  we propose an audio-textual cross-modal representation extractor to learn contextual representations directly from preceding speech. 
  Specifically, it consists of two modal-related encoders, extracting high-level latent features from speech and the corresponding text, and a cross-modal encoder, which aims to learn the correlation between speech and text. 
   We randomly mask some input tokens and input sequences of each modality. Then a token-missing or modal-missing prediction with a modal-level CTC loss on the cross-modal encoder is performed. Thus, the model captures not only the bi-directional context dependencies in a specific modality but also relationships between different modalities. 
  Then, during the training of the conversational ASR system, the extractor will be frozen to extract the textual representation of preceding speech, while such representation is used as context fed to the ASR decoder through attention mechanism. 
  The effectiveness of the proposed approach is validated on several Mandarin conversation corpora and the highest
  character error rate (CER) reduction up to 16\% is achieved on the MagicData dataset.
\end{abstract}

\section{Introduction}
  Conversational speech recognition is an important task in the field of automatic speech recognition (ASR), which specifically aims to transcribing spontaneous conversational speech into text~\cite{xiong2017toward}. 
  The leverage of conversation context has proven to be an effective way to optimize the performance of conversational ASR~\cite{kim2019cross}.
  Previous works usually 
  extract contextual information
  from transcripts of preceding speech in conversations, 
  such as long context language models~\cite{irie2019training,xiong2018session, mikolov2010recurrent,mikolov2012context},  
  reranking models~\cite{chiu2021cross} 
  and other methods~\cite{masumura2021hierarchical}.
  However, 
  at inference,
  hypotheses of the preceding utterances are used instead of ground truth transcripts to extract contextual representations.
  As a result,
  new errors may be introduced by the errors in historical ASR hypotheses when recognizing the current utterance.
  
  In order to avoid the error propagation caused by historical recognition errors, 
  a straightforward idea is to directly use previous speech as contextual features. 
  Some researchers use attention mechanism to transmit information in historical speech to current utterance~\cite{kim2018dialog, kim2019cross}, 
  or simply increase the length of input speech features~\cite{hori2020transformer, hori2021advanced}. 
  The above methods directly use low-level speech features (MFCC or Fbank) as additional historical dependencies, which may also bring in lots of redundant information, like environmental noises. Some researchers adopted the posterior probabilities as historical textual information. Although it may eliminate the noisy information, it is hard to learn effective representations with the limited labeled training data and could not take advantage of plenty of unlabeled data. Thus, it becomes urgent to extract useful context in the preceding speech while filtering out redundant features.
    Recently, the multimodal representation learning methods have also attracted wide attention. Multimodal methods can learn to capture long-distance dependencies between different modalities, such as voice, text, and image, leading to superior improvements on various downstream tasks~\cite{chung2021w2v, ao2021speecht5, liu2021opt, baevski2022data2vec}. Using cross-modal representation learning to extract linguistic representations from speech and eliminate the information that is not helpful for speech recognition is a feasible scheme for leveraging useful acoustic context in conversational ASR. A large scale of unlabeled data can be effectively adopted for the representation learning as well.
  
  In this paper, we propose a more effective way to better leverage acoustic context for conversational ASR. We introduce a cross-modal representation extractor that effectively takes the advances from pretrained speech and language models and subsequently provides acoustic context to a Conformer-based~\cite{gulati2020conformer} ASR.
  Specifically, it consists of two pre-trained \textit{modal-related encoders} -- Wav2Vec2.0~\cite{baevski2020wav2vec} and RoBERTa-wwm~\cite{bianchi-etal-2021-pre} -- extracting high-level latent features from speech and the corresponding text, and a \textit{cross-modal encoder}, which aims to learn the correlation between speech and text. 
  We randomly mask some input tokens and input sequences of each modality. Then a token-missing or modal-missing prediction with a modal-level CTC loss on the cross-modal encoder is performed. Thus, the model can capture not only the bi-directional context dependencies in a specific modality but also relationships between the two modalities. 
  In the training of the conversational ASR system, the extractor will be frozen to extract the textual representation of preceding speech, while such representation is used as context fed to the ASR decoder through attention mechanism.
  By this way, the acoustic context directly benefits the decoding process of the current speech utterance. 
  We verify the proposed approach on three Mandarin 
  conversational datasets -- HKUST~\cite{liu2006hkust}, Datatang and MagicData, and results show that the proposed method achieves 5\% -- 16\% character error rate (CER) reduction.

\begin{figure}[htbp]
\centering
\includegraphics[scale=0.5]{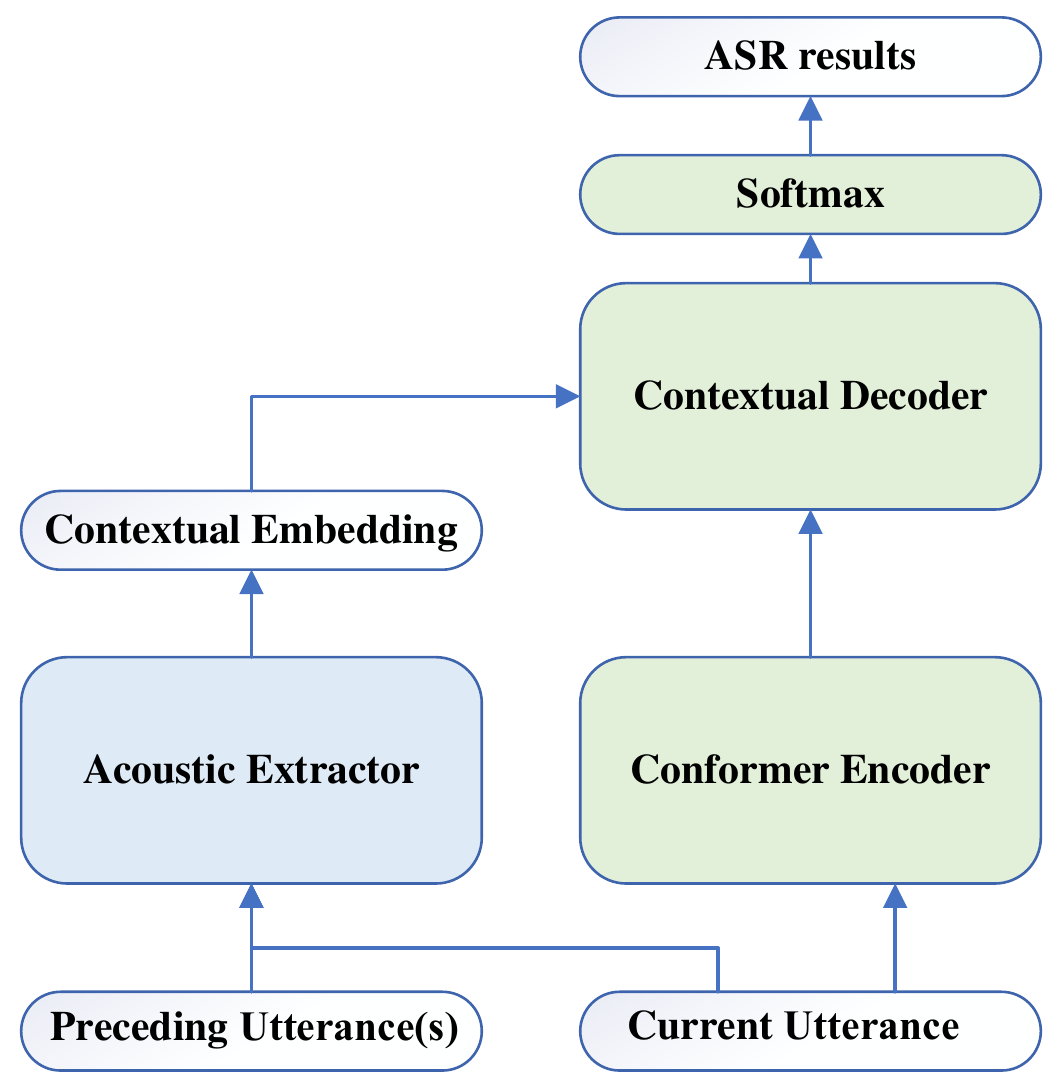}
\caption{
The diagram of the proposed method, the inputs are speech sequences.}
\label{fig:frame}
\end{figure}

\section{The Proposed Method}
Figure~\ref{fig:frame} shows the diagram of the proposed conversational ASR system. 
The proposed cross-modal representation extractor is used as a textual representation extractor to extract contextual information from speech.
The proposed conversational ASR system is trained in a two-stage way.
In the first stage,
the representation extractor is trained as shown in Figure~\ref{fig:ext}.
In Figure~\ref{fig:ext}, the context extractor is named Acoustic Extractor.
Audio and textual embeddings are derived from paired speech and transcripts using the speech encoder and the text encoder respectively. 
Then, these embeddings are sent to a cross-modal encoder to obtain the cross-modal representations.
The representation extractor learns correlations between paired speech and transcripts in different data granularities using multitask learning.
In the second stage,
the text encoder in the mutimodal representation extractor is discarded.
The extractor alternatively learns contextual representation from speech.
At both training and testing the ASR module,
the contextual representations are integrated into the decoder of the ASR module by attention mechanism.
The following subsections will describe each component in detail.

\subsection{Contextual Representation Extractor}
\subsubsection{Speech Encoder}
The speech encoder consists of a pre-trained speech representation model, Wav2vec2.0, and a linear layer.
The pre-trained model adopts the same architecture as Wav2vec2.0 large~\cite{baevski2020wav2vec} and is trained on WenetSpeech,
which contains more than 10000 hours annotated Mandarin speech~\cite{zhang2021wenetspeech}. 
The linear layer is to ensure that the outputs of the speech encoder and the text encoder have the same dimension. 

\subsubsection{Text Encoder}
We directly adopt the released pre-trained model, RoBERTa-wwm-ext~\cite{cui-etal-2021-pretrain}, as the text encoder,
which is trained with 5.4B words in-house text data including encyclopedia, news, and question answering webs.

\subsubsection{Cross Model Encoder}
The cross model encoder (CME) consists of three transformer blocks~\cite{vaswani2017attention}.
The speech embedding $\textbf{A}$ and text embedding $\textbf{T}$ obtained from the speech encoder and the text encoder respectively, 
are sent into the CME to obtain high-dimensional cross modal contextualized representations:
\begin{equation}
\textbf{H} = \rm{CME}(\textbf{A};\textbf{T}),
\end{equation}
where $(\cdot;\cdot)$ is the splicing operation.

\subsubsection{Training Objectives}
\textbf{Character-level loss.} 
We train the contextual extractor with character-level loss and modal-level loss. 
We mask 30\% of the text and speech embeddings respectively to get the masked feature representation $\textbf{A}_{\rm remain}$ and $\textbf{T}_{\rm remain}$. 

For the textual-modal character-level learning, 
similar to BERT, 
we predict the masked characters $\textbf{T}_{\rm masked}$ through 
the original speech $\textbf{A}$ and the remained parts of the transcript  $\textbf{T}_{\rm remain}$.
Specifically, 
we achieve this goal by minimizing the following negative logarithmic loss:
\begin{equation}
    L_{\rm MLM}(\theta)=-E_{(\textbf{A},\textbf{T})} {\rm log}P_{\theta}(\textbf{T}_{\rm masked}|\textbf{A},\textbf{T}_{\rm remain}),  
\end{equation}
where $\theta$ is the trainable parameters.

For the audio-modal character-level learning, 
in order to express the continuous context features of speech,
we refer to data2vec~\cite{baevski2022data2vec} to restore the masked speech information by minimizing the distance between the representation derived from the original speech $\textbf{A}$ and that from the masked speech $\textbf{A}_{\rm remain}$. 
Firstly, we calculate the latent representations:
\begin{equation}
\begin{aligned}
\textbf{H}_{\rm remain}&=\rm{CME}(\textbf{A}_{\rm remain};\textbf{T}), \\
\textbf{H}_{\rm other}&=\rm{CME}(\textbf{A}_{\rm other};\textbf{T}_{\rm other}),
\end{aligned}
\end{equation}
where $\textbf{A}_{\rm other}$ and $\textbf{T}_{\rm other}$ are paired speech and transcript randomly selected from other training sentences in the same batch of current sentence. 
The loss function is expressed as follows,
\begin{equation}
    L_{\rm MAM}(\theta)=-{\rm log}\frac{{\rm sim}(\textbf{H}, \textbf{H}_{\rm remain})}{{\rm sim}(\textbf{H}, \textbf{H}_{\rm remain})+{\rm sim}(\textbf{H}, \textbf{H}_{\rm other})},
\end{equation}
where ${\rm sim}(\cdot,\cdot)$ the Laplace distance.

\begin{figure}[htbp]
\centering
\includegraphics[scale=0.5]{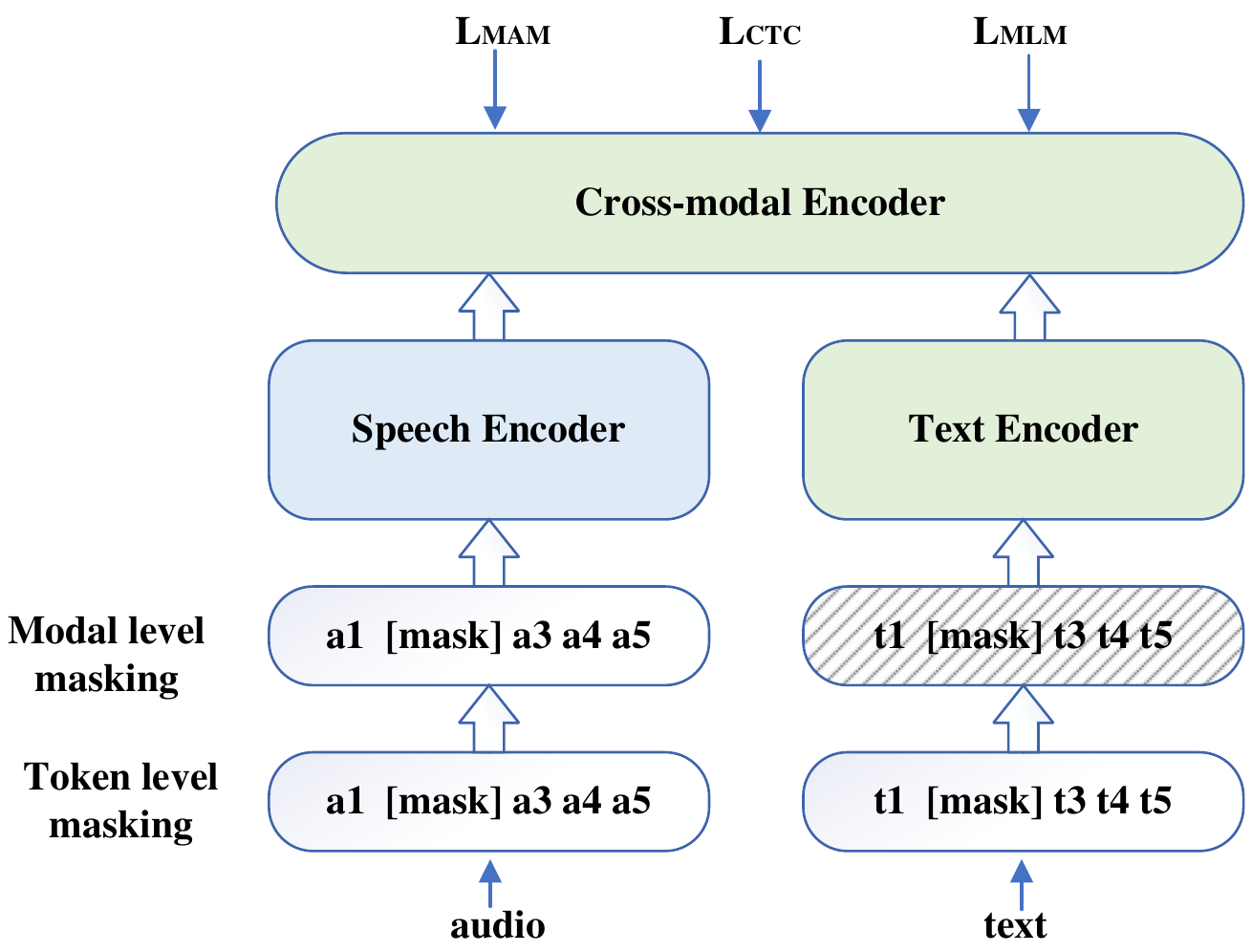}
\caption{
The diagram of the proposed contextual representation extractor. }
\label{fig:ext}
\end{figure}

\noindent\textbf{Modal-level loss.} 
In order to make the model learn the relationship between paired speech and transcript  in modal-level, inspired by~\cite{liu2021opt}, 
we randomly mask the original speech or text embeddings with the probability of 30\%, represented by the shaded part in figure~\ref{fig:ext}. Here, 
to better adapt to the downstream ASR task, 
we use the CTC loss to decode the current cross-modal representations into corresponding transcripts, 
so that the contextual representation extractor can learn more accurate alignment information between speech and transcript. 

\noindent\textbf{Final loss.} Finally, after integrating the above learning strategies, the final loss can be expressed as:
\begin{equation}
    L_{\rm pre} = {\alpha}L_{\rm CTC}+{\beta}L_{\rm MLM}+(1-\alpha-\beta)L_{\rm MAM},
\end{equation}
where $\alpha$ and $\beta$ control the effects of different losses.
During training, the parameters of Wav2vec2.0 and RoBERTa are frozen.

\subsection{Contextual Conformer ASR Model}
\subsubsection{Conformer Encoder}
Conformer~\cite{gulati2020conformer} organically combines convolutions with self-attention in ASR task. 
This structure learns the interaction of global information through self attention mechanism, and learns the representation of local features through convolution neural network, leading to superior performance. 
We stack conformer blocks as the encoder of the ASR model, each conformer block includes a multi-head self-attention layer (MHSA), a convolution layer (CONV) and a feed-forward layer (FFN). Assuming the input of the $n$-th block is $\textbf{z}_{n}$, operations in this block can be expressed as:
\begin{equation}
\label{Conformer}
\textbf{s}_{n} = \rm{MHSA}(\textbf{z}_{n}) + \textbf{z}_{n},
\end{equation}
\begin{equation}
\textbf{c}_{n} = \rm{CONV}(\textbf{s}_{n}),
\end{equation}
\begin{equation}
\textbf{z}_{n+1} = \rm{FFN}(\textbf{c}_{n}) + \textbf{c}_{n}.
\end{equation}

\subsubsection{Contextual Decoder}
A transformer with additional corss-attention layer is used as the decoder. 
For example, a conversation with K sentences, 
which speech and transcripts are $\{\textbf{A}_1,\textbf{A}_2,...\textbf{A}_i,\textbf{A}_{i+1},...,\textbf{A}_k,\}$ and 
$\{\textbf{T}_1,\textbf{T}_2,...\textbf{T}_i,\textbf{T}_{i+1},...,\textbf{T}_k,\}$.
We first generate the textual embeddings of current speech $\textbf{A}_i$ 
and previous speech.
We send the speech to be processed to the extractor together with a dummy embedding $\textbf{P}$ which is a zero vector.
The extraction method is as follows when only the previous one sentence is used,
\begin{equation}
    \textbf{H}_i=\rm{CME}(\textbf{A}_i;\textbf{P}), 
\end{equation}
\begin{equation}
    \textbf{H}_{i-1}=\rm{CME}(\textbf{A}_{i-1};\textbf{P}).
\end{equation}
Then we splice current textual embedding $\textbf{H}_i$
with the contextual embedding $\textbf{H}_{i-1}$ to obtain the final contextual embedding $\textbf{H}_{\rm context}=(\textbf{H}_{i-1};\textbf{H}_{i})$.

We feed same contextual embedding $\textbf{H}_{\rm context}$ into each block of the decoder to make the decoder learn the context information extracted by textual extractor. 
Assuming the output of the Conformer encoder is $\textbf{z}_i$ and
the input of the $m$-th layer in the decoder is $\textbf{t}_{m}$,
the operations in the $m$-th layer can be expressed as:
\begin{equation}
    \textbf{o}_{m} = \rm{MHSA}(\textbf{t}_{m}) + \textbf{t}_{m},
\end{equation}
\begin{equation}
    \textbf{r}_{m} = \rm{MHA}(\textbf{o}_{m},\textbf{z}_{i}) + \textbf{o}_{m},
\end{equation}
\begin{equation}
    \textbf{t}_{m+1} = \rm{MHA}(\textbf{r}_{m},\textbf{H}_{context}) + \textbf{r}_{m},
\end{equation}
here, MHA means the muti-head attention layer. The output of the last layer of the decoder is used to predict character probabilities through a softmax function.
\section{Experiments}

\subsection{Datasets}
We conduct experiments on three Mandarin conversation datasets--HKUST~\cite{liu2006hkust}, 
Datatang (DDT) 
and MagicData.
HKUST and DDT contain 200 hours and 350 hours speech data with the sampling rate of 8kHz respectively.
The dev sets in these datasets are used to evaluate the proposed method. 
The MagicData contains 160 hours of speech data with the sampling rate of 16kHz,
and the test set is used to evaluate the proposed method.  

\subsection{Implementation Details}
 We use 80-dimensional log-Melfilterbank (fbank) acoustic features as the input features of the Conformer encoder.
 The input of the textual representation extractor is the raw audio sequence. 
 Speed perturbation at ratio 0.9, 1.0, 1.1 with SpecAugment~\cite{park2019specaugment} is used to enhance the robustness of the ASR model. 
 In particular, in order to effectively utilize the Wav2vec 2.0 model,
 the speech data with the sampling rate of 8kHz are upsampled to 16KHz before being fed into the contextual representation extractor.
 
 The cross-modal extractor consists of 3 transformer blocks. 
 When training the ASR model, 
 the parameters in the acoustic contextual representation extractor are frozen. 
 For each corpus, 
 the configurations of acoustic features and the Conformer model are almost the same as the ESPnet Conformer recipes~\cite{watanabe2018espnet}.
 The Conformer encoder has one Conv2D module as the downsampling module 
 and is followed with 12 Conformer blocks.
 Each block has a multi-head attention with 8 heads with 256 units and 
 a feed-forward layer with 2048 units. 
 The contextual decoder has 6 transformer blocks and an embedding layer. 
 
 We train the baseline models using independent vocabulary from each dataset. 
 The baseline models have 3653, 3126 and 4048 output units (characters) for HKUST, DDT and MagicData. 
 For the textual representation extractor model, 
 all datasets share RoBERTa-wwm's vocabulary, 
 which has 21128 characters. 
 
 Instead of training from scratch, 
 we train the contextual ASR model by fine-tuning the baseline conformer ASR model,
 which leads to faster and smoother convergence.
 In order to do a fair comparison, 
 the baseline model will continue to train the same number of epochs.

 In order to verify the effectiveness of our method, we reproduce the results of conversational speech recognition based on text context~\cite{kun2022conversational}, as shown in the penultimate row of Table ~\ref{tab:all_results}. Additionally, three-layer transformer language models with 2048 units using the transcripts of each dataset is trained for experiments. 
 All experiments are done with the ESPnet~\cite{watanabe2018espnet} toolkit. 
\subsection{Results and Analysis}
Table~\ref{tab:all_results} shows the results of our proposed method. 
The proposed contextual ASR model 
achieves up to 16\% relative CER reduction compared with the baseline in row 1, 
as well as outperforms the vanilla Conformer model~\cite{guo2021recent} and the CVAE-Conformer model~\cite{kun2022conversational}.
\begin{table*}
\centering
\caption{CER comparation of different end-to-end models on three Mandarin datasets. The AcousticCur means using the textual embedding of current sentence, AcousticCon means the model using the textual embeddings of current sentences and previous sentence, ExtLM represents that additional language models are used in ASR decoding.}
\label{tab:all_results}
\begin{tabular}{lcccccccc}
\toprule
\textbf{Method} &\textbf{AcousticCur} & \textbf{AcousticCon} & \textbf{ExtLM}& \textbf{HKUST} & \textbf{DDT} & \textbf{MagicData} \\
\midrule
\multirow{6}{*}{\textbf{Acoustic Contextual Conformer}}   
& - & - &-       & 20.2 & 20.6 & 18.6 \\
& - & - &Y       & 20.1 & 20.4 & 18.5 \\ 
& Y & - &-       & 19.5 & 19.3 & 15.9 \\
& Y & - &Y       & 19.3 & 19.4 & 15.8 \\
& - & Y &-       & \textbf{19.1}        & \textbf{18.8}          & \textbf{15.5}           \\ 
& - & Y &Y       & 19.2 & 18.9 & 16.0 \\ \hline
\textbf{CVAE-Transformer}~\cite{kun2022conversational} &-&-&- &20.1  &20.0 &  17.6& \\ 
\textbf{ESPnet Conformer}~\cite{guo2021recent} &-&-&- &22.2  & & & \\ \bottomrule
\end{tabular}
\end{table*}


%

\subsubsection{Effect of Acoustic Contextual Representation} 
By comparing the experimental results in rows 1, 3 and 5 in Table 1, 
we can find that the proposed method can improve the recognition accuracy even if we only extract the contextual representation of the current speech utterance $\textbf{A}_i$. After adding the contextual representation of the previous speech utterance $\textbf{A}_{i-1}$, that is, when using the acoustic context representation of $\textbf{A}_i$ and $\textbf{A}_{i-1}$ at the same time, the performance of our recognition system has been further improved.

To verify whether the proposed acoustic contextual extractor can learn textual information from speech, 
we compared the proposed method with external language models. 
From the results in rows 1 and 2 in Table~\ref{tab:all_results}, we can find that 
transformer language models can improve the recognition accuracy of the baseline ASR model.
However, as shown in rows 3 and 4 in Table~\ref{tab:all_results}, 
the transformer language models do not achieve any improvement after adopting the proposed method. 
Similar results are shown in rows 5 and 6,
when we additionally integrate the contextual representation of the previous sentence.
It shows that our extractor obtains abundant textual information, after introducing contextual representation, the ASR model can  
reduce the dependence on external language models.
\begin{table}[hbpt]
\caption{CER comparison to Wav2vec2.0 pretrained model. All results are without using language models.}
\label{tab:wav2vec}
\begin{tabular}{p{3cm}lll}
\toprule                                       & \textbf{HKUST} & \textbf{DDT}    &   \textbf{MagicData}\\ \midrule
baseline                                      & 20.2  & 20.6   &    18.6   \\ 
Wav2vec2.0 pretrain                           & 20.1  & 20.4   &    18.2   \\ 
AcousticCon (Prop.)                           & \textbf{19.5}  & \textbf{19.3}   &    \textbf{15.9}    \\ \bottomrule

\end{tabular}
\end{table}
\subsubsection{The Contexual Information of Wav2vec2.0}
The pre-trained speech model Wav2vec2.0 contains abundant speech context information from its rich unsupervised training corpus and self supervised learning strategies. 
Our purpose is to make full use of the effective information in the pretrained model using our method. But at the same time, we want to avoid relying too much on the the pretrained model itself instead of learning the cross-modal information between speech and text. 

We design a speech recognition model with Wav2vec2.0 for comparison. Specifically, we use the same pre-trained Wav2vec2.0 model as the encoder of the speech recognition model. At training, we freeze the parameters in the pre-trained model for a certain number of steps before training all the parameters together. 

As shown in Table~\ref{tab:wav2vec},
although the pretrained model can improve the accuracy of recognition, our method achieves significantly better results. 
This shows that our model not only makes use of the representation ability of the pretrained model, but also effectively obtains the cross-modal textual representation.

\subsubsection{The Length and Location of Historical Information} 
The length and location of historical speech utterances used to extract contextual representations usually affect the performance of ASR models.
We study the affects of history length and locations on HKUST and MagicData sets. 
Assuming that the current speech to be recognized is $\textbf{A}_i$ (cur), $\rm{AcousticCon}_{one}$ represents the previous sentence $\textbf{A}_{i-1}$ and $\rm{AcousticCon}_{two}$ represents the previous sentence $\textbf{A}_{i-2}$.

From Table \ref{tab:length}, we can find that the closer the sentence is to the current sentence, the more helpful it is to improve the recognition accuracy of the current sentence. However, inputting the textual features of the previous two sentences at the same time does not achieve better results, which may be caused by the decoder's inability to learn appropriate concerns from the long historical information.

\begin{table}[htbp]
\centering
\caption{Comparison of length and location using historical information. $\rm{AcousticCon}_{one}$ means using $\rm{AcousticCon}$ with previous one sentence, $\rm{AcousticCon}_{two}$ means use the penultimate sentence.}
\label{tab:length}
\begin{tabular}{p{3.5cm}ll}
\toprule                                                     & \textbf{HKUST} & \textbf{MagicData} \\ \midrule
$\rm{AcousticCur}$                                    & 19.5    & 15.9       \\ 
$\rm{AcousticCon}_{one}$                        & \textbf{19.1}  & \textbf{15.5}       \\ 
$\rm{AcousticCon}_{two}$                        & 19.3     &      15.8  \\ 
$\rm{AcousticCon}_{one+two}$                 & 19.3          &  15.9            \\ \bottomrule
\end{tabular}
\end{table}
\section{Conclusions}
In this paper, we propose a cross-modal representation extractor to learn contextual information from speech, and use the representation for conversational ASR through the attention mechanism. 
The cross-modal representation extractor consists of two pretrained single-modal encoder, Wav2vec2.0 and RoBERTa, and a cross-modal encoder.
The textual representation extracted from current and previous speech is sent to the decoder module of ASR, which reduces the relative CER by up to 16\% on data sets MagicData, DDT and HKUST. 
In the future work, we will explore the impact of different pretrained speech models and language models on the extractor, as well as more effective ways to integrating the contextual representation into conversational speech recognition.
\bibliographystyle{IEEEtran}

\bibliography{mybib}


\end{document}